\documentclass[sn-mathphys]{sn-jnl}

\usepackage{siunitx}
\usepackage[version=3]{mhchem}
\usepackage{xcolor}
\usepackage[superscript,biblabel]{cite}
\usepackage[pagewise]{lineno}

\jyear{2023}

\theoremstyle{thmstyleone}

\theoremstyle{thmstyletwo}

\theoremstyle{thmstylethree}

\DeclareSIUnit\year{yr}
\DeclareSIUnit{\wtpercent}{wt\%}
\DeclareSIUnit\molar{M}

\begin{document}
\title[Isotopic constraints on lightning]{Isotopic constraints on lightning as a source of fixed nitrogen in Earth's early biosphere}

\author*[1,2,3,4]{\fnm{Patrick} \sur{Barth}}\email{pb94@st-andrews.ac.uk}
\author[1,2]{\fnm{Eva E.} \sur{Stüeken}}
\author[4,5]{\fnm{Christiane} \sur{Helling}}
\author[1,3,6]{\fnm{Lukas} \sur{Rossmanith}}
\author[2]{\fnm{Yuqian} \sur{Peng}}
\author[7,8]{\fnm{Wendell} \sur{Walters}}
\author[1,2]{\fnm{Mark} \sur{Claire}}

\affil*[1]{\orgdiv{Centre for Exoplanet Science}, \orgname{University of St Andrews}, \orgaddress{\street{North Haugh}, \city{St Andrews}, \postcode{KY16 9SS}, \country{UK}}}

\affil[2]{\orgdiv{School of Earth \& Environmental Sciences}, \orgname{University of St Andrews}, \orgaddress{\street{Bute Building, Queen's Terrace}, \city{St Andrews}, \postcode{KY16 9TS}, \country{UK}}}

\affil[3]{\orgdiv{SUPA, School of Physics \& Astronomy}, \orgname{University of St Andrews}, \orgaddress{\street{North Haugh}, \city{St Andrews}, \postcode{KY16 9SS}, \country{UK}}}

\affil[4]{\orgdiv{Space Research Institute}, \orgname{Austrian Academy of Sciences}, \orgaddress{\street{Schmiedlstrasse 6}, \city{Graz}, \postcode{A-8042}, \country{Austria}}}

\affil[5]{\orgdiv{Fakultät für Mathematik, Physik und Geodäsie}, \orgname{TU Graz}, \orgaddress{\street{Petersgasse 16}, \city{Graz}, \postcode{A-8010}, \country{Austria}}}

\affil[6]{\orgdiv{School of Chemistry}, \orgname{University of St Andrews}, \orgaddress{\street{North Haugh}, \city{St Andrews}, \postcode{KY16 9ST}, \country{UK}}}

\affil[7]{\orgdiv{Institute at Brown for Environment and Society}, \orgname{Brown University}, \orgaddress{\street{85 Waterman St}, \city{Providence}, \postcode{02912}, \state{Rhode Island}, \country{USA}}}

\affil[8]{\orgdiv{Department of Earth, Environmental and Planetary Sciences}, \orgname{Brown University}, \orgaddress{\street{324 Brook St}, \city{Providence}, \postcode{02912}, \state{Rhode Island}, \country{USA}}}

\abstract{
Bioavailable nitrogen is thought to be a requirement for the origin and sustenance of life. 
Before the onset of biological nitrogen fixation, abiotic pathways to fix atmospheric \ce{N2} must have been prominent to provide bioavailable nitrogen to Earth’s earliest ecosystems. 
Lightning has been shown to produce fixed nitrogen as nitrite and nitrate in both modern atmospheres dominated by \ce{N2} and \ce{O2} and atmospheres dominated by \ce{N2} and \ce{CO2} analogous to the Archaean Earth. 
However, a better understanding of the isotopic fingerprints of lightning-generated fixed nitrogen is needed to assess the role of this process on the early Earth. 
Here, we present results from spark discharge experiments in \ce{N2-CO2} and \ce{N2-O2} gas mixtures. 
Our experiments suggest that lightning-driven nitrogen fixation may have been similarly efficient in the Archaean atmosphere, compared to modern times. 
Measurements of the isotopic ratio ($\delta ^{15}$N) of the discharge-produced nitrite and nitrate in solution show very low values of $-6$ to $-15 \text{\textperthousand}$ after equilibration with the gas phase with a calculated endmember composition of $-17 \text{\textperthousand}$. 
These results are much lower than most $\delta ^{15}$N values documented from the sedimentary rock record, which supports the development of biological nitrogen fixation earlier than $\SI{3.2}{Ga}$. 
However, some Paleoarchean records ($\SI{3.7}{Ga}$) may be consistent with lightning-derived nitrogen input, highlighting the potential role of this process for the earliest ecosystems.
}


\maketitle

On Earth, nitrogen is an essential building block for biological macromolecules such as DNA, RNA, and proteins and, therefore, must have been available for the origin of life and for the sustenance of early ecosystems.
The most abundant form of nitrogen at the Earth's surface is atmospheric \ce{N2} gas; however, this molecule is relatively inert, requiring dedicated nitrogenase enzymes or high energy to be converted into more bioavailable forms (nitrogen fixation). 
On the modern Earth, more than 97\% of pre-industrial \ce{N2} fixation is carried out by microorganisms; only a few percent of fixed nitrogen are produced by abiotic sources, the main source being lightning\cite{Postgate1998,Fowler2013}. 
On the early Earth, before the origin of life and the onset of biological \ce{N2} fixation, these abiotic sources such as lightning must have been the dominant producer of bioavailable nitrogen. 
Lightning can occur when in a cloudy atmosphere cloud particles carry excess charges over long distances, building up large electric fields that eventually produce a lightning strike\cite{Beasley1982}.
Previous studies have shown that spark discharges produce nitrogen oxides in today's atmosphere \cite{Cavendish1788,Yung1979} as well as in \ce{N2}-\ce{CO2}-dominated atmospheres\cite{NnaMvondo2001,Navarro-Gonzalez2001,Summers2007}, and models suggest that a significant flux of these lightning products to the Earth's surface could potentially have fuelled prebiotic chemistry \cite{Kasting1981,Wong2017,Adams2021}.
However, it has so far not been possible to verify these model predictions and the role of lightning in the evolution of life, because the isotopic fingerprint of this nitrogen source was unknown, i.e. the ratio of the nitrogen isotopes \ce{^{15}N} and \ce{^{14}N} expressed as $\delta\ce{^{15}N} = [(\ce{^{15}N}/\ce{^{14}N} )_\mathrm{sample} / (\ce{^{15}N}/\ce{^{14}N})_\mathrm{standard} - 1] \times 1000 \; \text{\textperthousand}$, where the standard is modern air. 
In one laboratory experiment, values near $0 \text{\textperthousand}$ were reported from the gas phase \cite{Hoering1957}. 
However, natural measurements of atmospheric nitrate from different sources suggested values of $-5 \text{\textperthousand}$ to $-15 \text{\textperthousand}$ \cite{Moore1977} with extreme values between $<-30 \text{\textperthousand}$ (nitrate and \ce{NO_{$x$}} emission from the Antarctic snowpack) and $>+5 \text{\textperthousand}$ (fossil fuel emissions)\cite{Shi2021}, indicating the wide range of sources of nitrate emissions on the modern Earth.
Because experimental measurements of the aquatic phase have not yet been conducted, it has so far been impossible to detect evidence of lightning from the sedimentary rock record of nitrogen isotope ratios\cite{Stueken2016,Stueken2021,Yang2019} (Fig.~1).

\begin{figure}[ht]
    \centering
    \includegraphics[width=0.8\columnwidth]{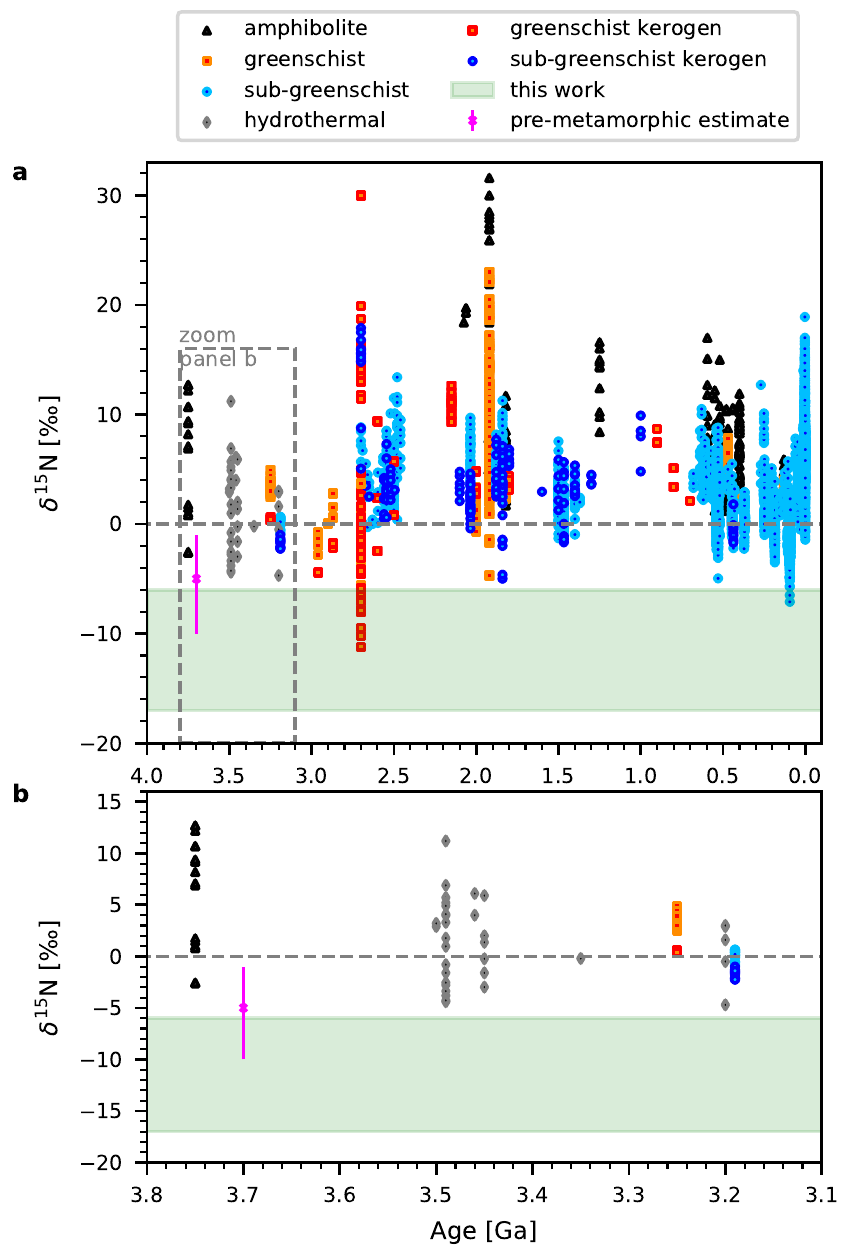}
    \caption{ 
    \textbf{Measurements of $\delta ^{15}$N in sedimentary rocks over geologic time, separated by metamorphic grade.}
    \textbf{a}, Data from Stüeken et al. \cite{Stueken2016,Stueken2021} and Yang et al. \cite{Yang2019}.
    Errors smaller than symbols. 
    \textit{Pink x:} pre-metamorphic estimate of 3.7 Ga old sample set ($-5 \text{\textperthousand}$, values between $-1$ and $-10 \text{\textperthousand}$ possible).
    \cite{Stueken2021}.
    Our results after letting gas and water equilibrate are indicated by green shading.
    \textbf{b}, Zoom into area indicated in panel a by grey box.}
    \label{Fig_d15N_Rock}%
\end{figure}

\begin{figure}[ht]
    \centering
    \includegraphics[width=0.8\columnwidth]{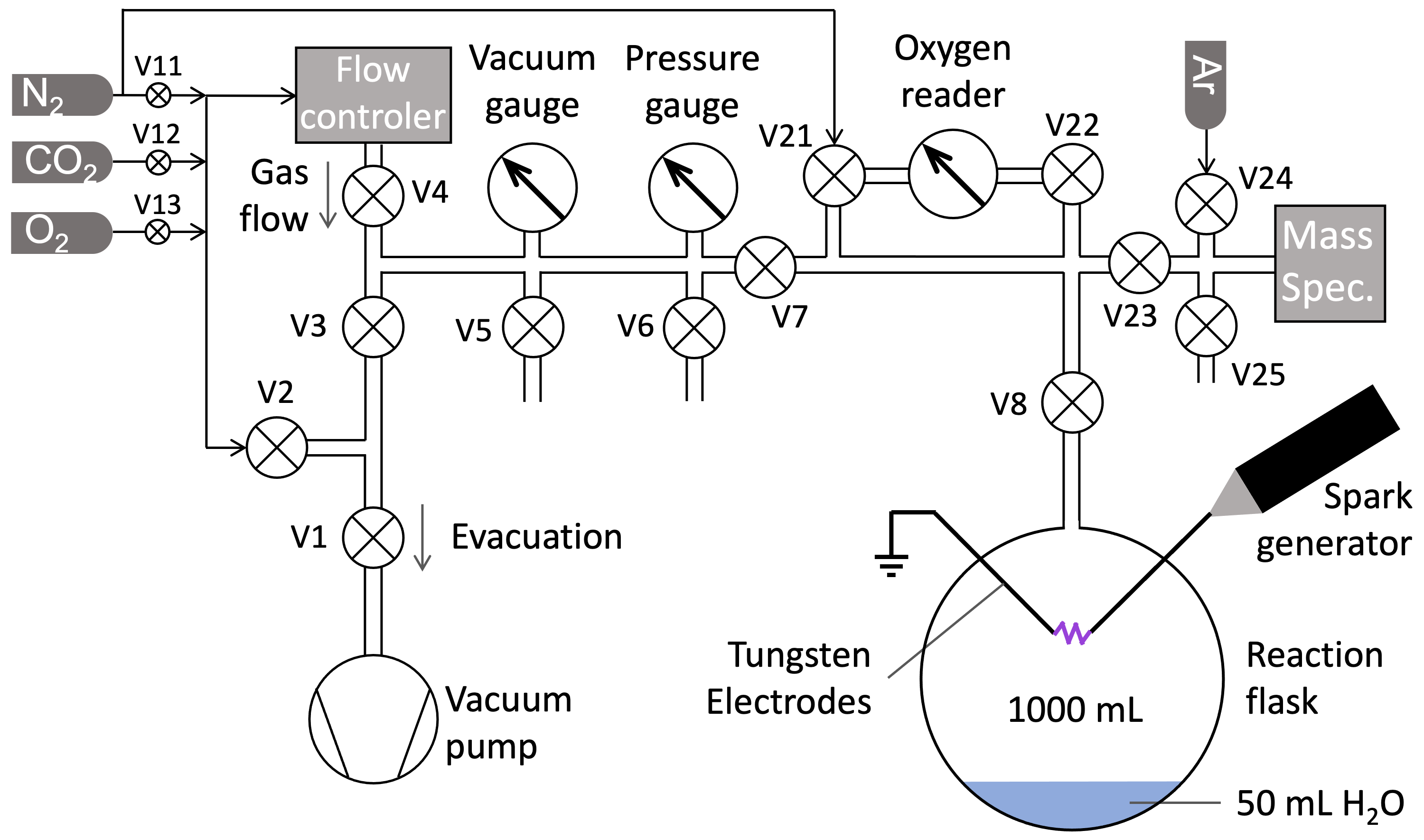}
    \caption{\textbf{Schematic of experimental setup of the discharge experiment.}
    See Methods for a detailed description of the procedure.}
    \label{Fig_ExpSetup}%
\end{figure}

In this study we identified an isotopic fingerprint with spark discharge experiments in \ce{N2-O2} and \ce{N2-CO2} gas mixtures. 
We know from measurements of fluid inclusions that the isotopic composition of \ce{N2} in early Earth's atmosphere was very similar to today's (within $\sim 2 - 3 \text{\textperthousand}$)\cite{Marty2013}, enabling us to compare our fingerprint to sedimentary rock samples.
The experimental setup resembled a Miller-Urey apparatus \cite{Parker2014} with a $\SI{1}{\litre}$ glass flask and two needle electrodes at a distance of $\sim \SI{1.5}{\centi\metre}$ (Fig.~2). 
A current (I) of $\SI{1}{\milli\ampere}$ at a voltage (U) of $\SI{49}{\kilo\volt}$ was applied continuously for a time span (t) of $15-\SI{60}{\minute}$ per experiment. 
The flask was filled with artificial gas mixtures to $\SI{1}{\bar}$, mimicking modern (85\% \ce{N2}, 14\% \ce{O2}), \ce{O2}- and \ce{CO2}-depleted (99.5\% \ce{N2}, 0.06\% \ce{O2}, $<$0.02\% \ce{CO2}), and Archean-like (83\% \ce{N2}, 16\% \ce{CO2}) atmospheres.
For the \ce{CO2} fraction in the Archean-like experiments we followed estimates predicting a \ce{CO2} mole fraction of 15 to 20\% in the early Archean\cite{Catling2020}.
In each case, $\SI{50}{\milli\litre}$ of deionized water at the bottom of the flask were used to trap soluble nitrogen oxides. 
Electrode spacing, energy input, experiment duration, fluid salinity and pH were varied and their effect on the results explored (see supplementary material).
We measured the isotopic composition and abundances of these dissolved oxides by gas-source mass spectrometry, and the composition of the gas before and after the experiment using an on-line quadrupole gas analyser (see Methods). 
The molecular concentrations allowed us to determine energy yields (molecules/Joule), where energy input was calculated as $E = 1/2 \; U I t$. 
The energy yield could in turn be used to estimate global fluxes to the Earth's surface for each of the major products.
The isotopic composition allows us to test if lightning might have been an important nutrient source for early life.

\section*{Energy yield of nitrogen fixation}

\begin{figure}[ht]
    \centering
    \includegraphics[width=0.7\columnwidth]{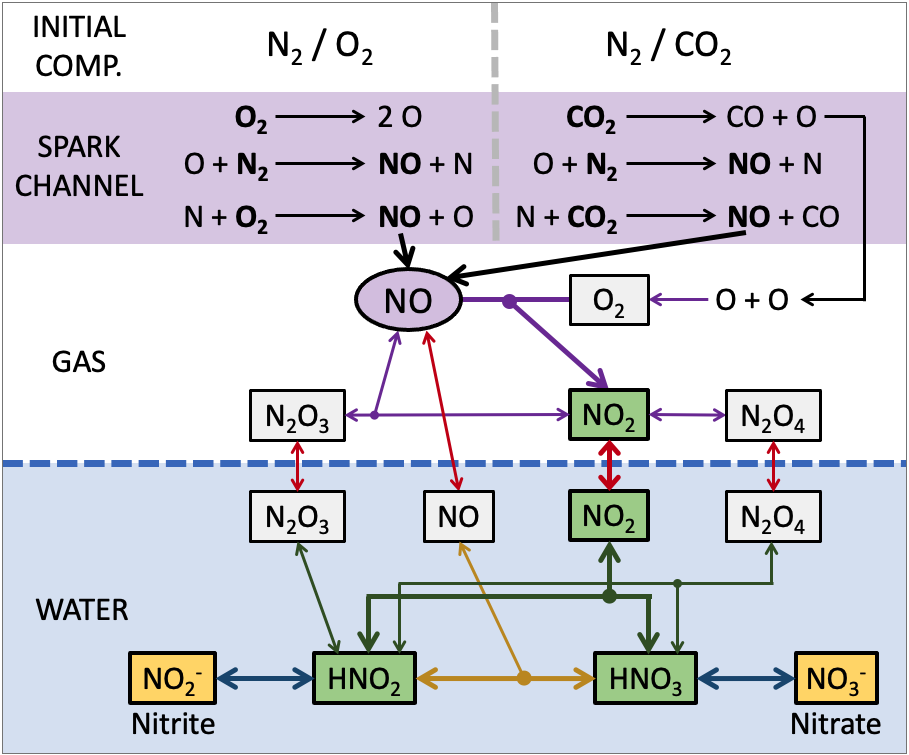}
    \caption{
    \textbf{Chemical pathways during spark discharge in \ce{N2-O2} (left) and \ce{N2-CO2} (right) gas mixtures.} 
    Once NO is produced, further oxidation follows the same reactions in both cases: NO is oxidised to \ce{NO2}, some being converted to \ce{N2O3} and \ce{N2O4}. 
    In \ce{N2-CO2} experiments, \ce{O2} is produced by the recombination of atomic oxygen.
    All gaseous nitrogen oxides equilibrate with water and convert to \ce{HNO2} and \ce{HNO3}. 
    \ce{HNO2} is thermodynamically unstable and will oxidise to \ce{HNO3}. 
    Both acids may dissociate to nitrite (\ce{NO2^-}) and nitrate (\ce{NO3^-}), respectively, depending on solution pH.
    In a low-pH environment, \ce{HNO2} is more abundant than \ce{NO2^-}, allowing further oxidation to \ce{HNO3} and nitrate. 
    The dominant pathways are indicated with bold arrows and coloured molecule labels.
    }
    \label{Fig_Lightning_Chem}
\end{figure}

\begin{figure}[!ht]
    \centering
    \includegraphics[width=0.8\columnwidth]{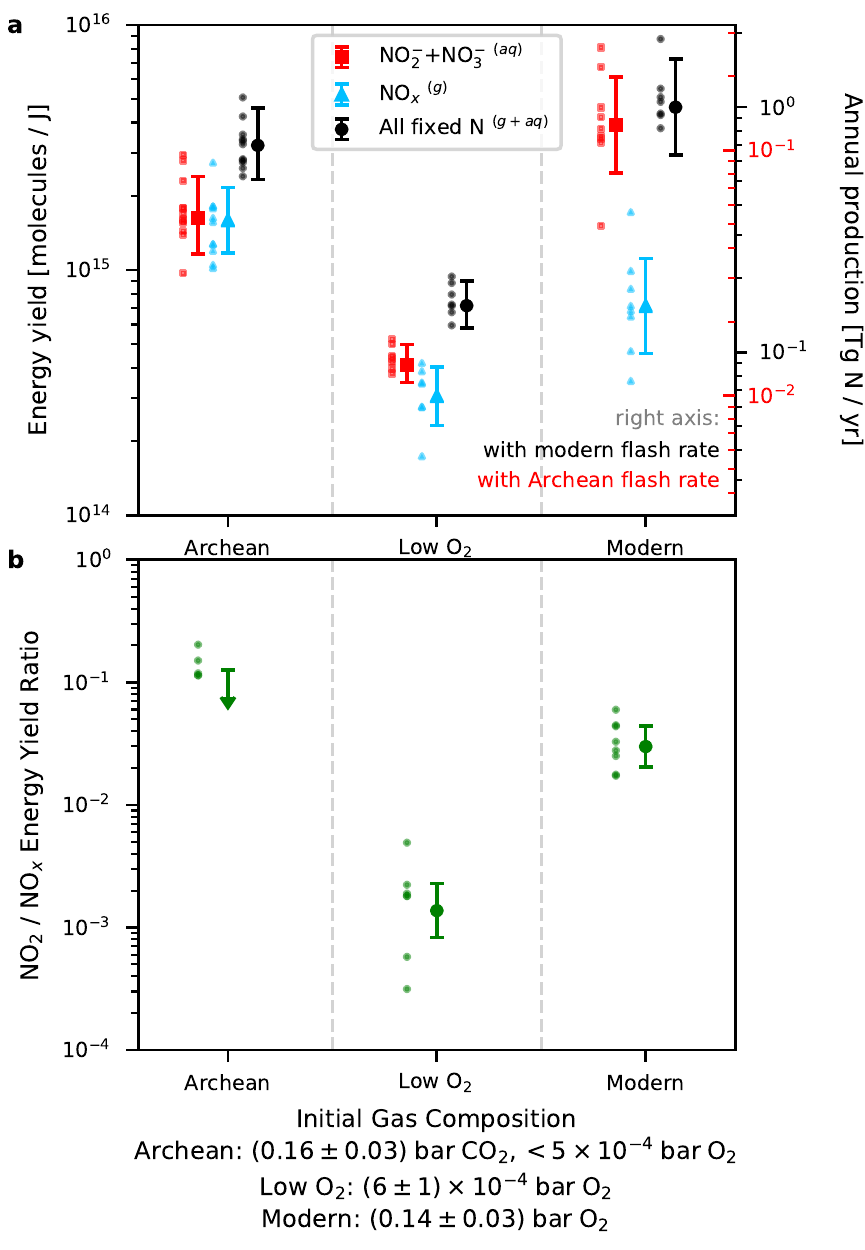}
    \caption{
    \textbf{Energy yield of fixed nitrogen products.}
    \textbf{a}, Combined gaseous products (\ce{NO_{$x$}}, blue), combined aqueous nitrate and nitrite (red), and total fixed nitrogen yield (black) for different initial gas compositions (difference to $\SI{1}{\bar}$ is \ce{N2}). 
    Individual measurements shown as small symbols, the large symbols represent the mean values (with SD error).
    Nitrite and nitrate are not separated, because nitrite conversion to nitrate was found to continue after the experiments.
    Right axis shows corresponding annual production for modern\cite{Christian2003} (black) and potential Archean\cite{Wong2017} (red) lightning flash rates in Tg of fixed nitrogen per year.
    \textbf{b}, Ratio of the energy yields of \ce{NO2} and \ce{NO_$x$} for different initial gas compositions of the experiment (same as above). 
    The data point for \ce{N2-CO2} gas represents only an upper limit due to interferences between \ce{NO2} and \ce{CO2} in the gas analyser.
    }
    \label{Fig_NFixation}
\end{figure}

The main nitrogen product that is expected to form during spark discharges in \ce{N2-O2} and \ce{N2-CO2} gas mixtures is NO \cite{Kasting1981,Miller1987}, which reacts further with \ce{O2} to produce \ce{NO2} and minor oxides (\ce{N2O3}, \ce{N2O4}) (Fig.~3). 
In the anoxic experiments, the \ce{O2} is provided by the recombination of atomic O after the dissociation of \ce{CO2}\cite{NnaMvondo2001}.
\ce{NO2} dissolves into the water and converts to aqueous \ce{HNO2} and \ce{HNO3}, which in turn dissociate to \ce{NO2^-} and \ce{NO3^-}, respectively. 
Figure~4a shows the energy yield for the major products in our experiments for the different gas compositions. 
Yields for dissolved ammonium were also measured but found to be very low ($<0.5\%$ of total fixed N) and are therefore not considered further, though our ammonium detection is consistent with previous observations of ammonium production in anoxic \ce{N2-CO2} spark experiments \cite{Cleaves2008}.
The modern and Archean-like scenarios both show a total energy yield for the sum of all fixed nitrogen species of $(3-5) \times 10^{15}$~molecules/J. 
Total yields are lower in the low-\ce{O2} experiments, likely because NO formation was oxygen-limited. 
The speciation of the fixed nitrogen was found to vary with atmospheric composition: 
In the modern experiments (14\% \ce{O2}), $\sim$85\% of the fixed nitrogen is stored in dissolved nitrate and nitrite. 
In the Archean-like experiments (16\% \ce{CO2}), only about 50\% is stored in nitrate and nitrite. 
The low-\ce{O2} scenario yielded a smaller fraction of aqueous nitrogen species, nearly half of the fixed nitrogen remained in the gas phase. 
This suggests that in the Archean-like experiments the \ce{CO2} provided enough oxygen for equally efficient NO formation as in an oxygen-rich atmosphere. 
However, \ce{O2} availability appears to limit the efficient conversion of NO to \ce{NO2} and aqueous species. 
This interpretation is supported by the low \ce{NO2}/NO ratios measured in the low-\ce{O2} experiment compared to the modern gas mixture (Figure~4b). 
Reliable \ce{NO2} measurements for the Archean-like experiments could not be obtained due to interference between \ce{NO2} and \ce{CO2} isotopologues at mass 46, but the relatively low abundance of aqueous compared to gaseous products suggests that the Archean-like \ce{NO2}/NO ratio was also lower, due to oxygen-limitation.

To estimate global fluxes of fixed nitrogen to the Earth's surface, based on our results, we assumed a lightning flash rate in Earth's atmosphere\cite{Christian2003} of $44\pm5 \; \si{\per\second}$ and for the average energy dissipated by one lightning flash \cite{Price1997} of $\SI{6.7}{\giga\joule}$.
Based on these assumption, our Archean-like and modern experiments predict the production of $0.7-\SI{1}{\tera\gram}$ fixed nitrogen per year. 
A global lightning flash rate of only $\SI{6.6}{\per\second}$\cite{Wong2017}, as proposed for the Archean \cite{Romps2014} would lower the annual fixed nitrogen production by lightning accordingly.
The results are lower for the low-\ce{O2} case (Fig.~4). 
Overall, these estimates are comparable to those of previous studies, including experiments that more closely mimic natural lightning conditions in terms of total energy input and spark length \cite{Wang1998,Cook2000,Schumann2007}, supporting our approach of simulating lightning chemistry with small spark experiments. 
Previous experiments with a laser-induced plasma to simulate lightning discharge \cite{NnaMvondo2001,Navarro-Gonzalez2001} found a 4-times lower NO energy yield (compared to our total fixed N yield) of $7 \times 10^{14} \; \si{molecules\per\joule}$ in \ce{N2-CO2} gas mixtures and a 20-times higher yield in modern air, compared to our measurements. 
However, it is uncertain how well a laser can mimic lightning chemistry. 
In a natural lightning channel, gas is heated within a few microseconds to a peak temperature of $\SI{30000}{\kelvin}$ \cite{Uman2003a,An2019}. 
The gas cools to $\sim \SI{3000}{\kelvin}$ over a few to tens of milliseconds \cite{Uman1968,Picone1981}. 
While the gas is cooling NO forms.
However, the timescales for thermochemical equilibrium to establish increases with further cooling from microseconds at $\SI{4000}{\kelvin}$ to $\SI{1000}{\year}$ at $\SI{1000}{\kelvin}$ \cite{Chameides1986}, causing the NO concentration to be fixed once the cooling time-scale drops below the equilibrium time-scale \cite{Zeldovich1966}. 
Hence, even though in our experiments the spark channel does not reach the temperature of a fully-developed lightning channel, NO was still observed to be the dominant product, which allows us to mimic natural atmospheric chemistry induced by lightning.
To further verify the validity of our results for natural settings, we ran a subset of experiments with artificial seawater and \ce{CaCO3} as a pH buffer (see supplementary material), and found no effect on the nitrite concentrations, indicating that our results are valid for both fresh and saline water. 
Experiments with anoxic seawater and $\SI{0.8}{\milli\molar}$ ferrous iron, akin to the Archean ocean\cite{Tosca2019}, had similar yields, suggesting that potential reactions between nitrogen oxides and ferrous iron\cite{Summers1993} are relatively minor and do not impact our conclusions.

\section*{Isotope fractionation during spark experiment}

\begin{figure}[!ht]
    \centering
    \includegraphics[width=0.8\columnwidth]{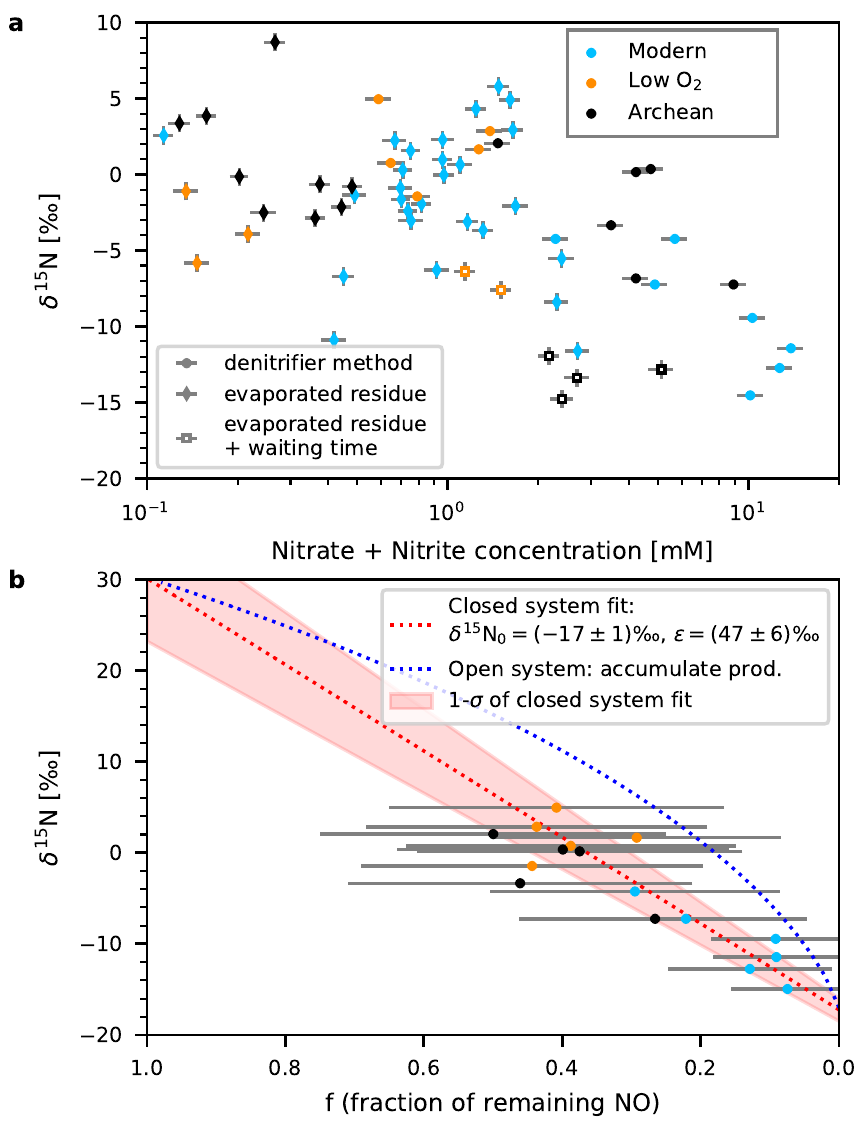}
    \caption{
    \textbf{Nitrogen isotope ratios of aqueous nitrate and nitrite.}
    \textbf{a}, 
    $\delta ^{15}$N against combined concentration of nitrate and nitrite in mM ($10^{-3}\si{mol\per\litre}$). 
    Data from three sets of experiments (symbol shape) and three gas mixtures (symbol color).
    Data points are individual measurements with uncertainties in concentration and $\delta ^{15}$N (errors in $\delta ^{15}$N for denitrifier method smaller than symbols). 
    \textbf{b}, $\delta ^{15}$N against remaining fraction of \ce{NO} $f = $[NO]$/($[NO]+[\ce{NO2}]+[\ce{NO2^-}]+[\ce{NO3^-}]$)$. \ce{NO} abundances to calculate $f$ only available for experiments analysed with the denitrifier method (circles). 
    Dashed red line is fit to a closed system Rayleigh fractionation ($\delta ^{15} \ce{N} (f) = \delta ^{15} \ce{N}_0 + \epsilon f$) with 1-$\sigma$ envelope, allowing to extrapolate to endmember where all NO oxidised to nitrate and nitrite.
    Blue line represents the $\delta ^{15}$N of the accumulated product in a Rayleigh fractionation assuming an open system with same $\delta ^{15} \ce{N}_0$ and $\epsilon$ as closed system.
    See text for discussion of open and closed system fractionation.
    Data points represent individual measurements with error in $f$ propagated from uncertainties of the measurements of the individual concentrations (NO, \ce{NO2}, \ce{NO2^-}, \ce{NO3^-}).}
    \label{Fig_d15N_Results}
\end{figure}

We analysed the nitrogen isotopic composition in the dissolved oxides with two different techniques (see Methods). 
For most experiments, the solutions were mixed with KOH powder, freeze-dried, and then analysed as dried salt. 
For a subset of experiments, dissolved nitrate and nitrite were treated with the denitrifier method and analysed as \ce{N2O} gas\cite{Sigman2001,Casciotti2002} which allowed separating out other nitrogen compounds such as ammonium. 
The two datasets combined define an overlapping trend of progressively more negative $\delta ^{15}$N values with higher combined nitrate and nitrite concentrations, with values as low as $-15\text{\textperthousand}$ (Fig.~5a).
Neither of the two methods was able to separate nitrite from nitrate, which is why we report them together.

Combined with the gas data, we find that the isotopic composition decreases as the fraction of NO remaining in the gas phase $f = [\ce{NO}]/([\ce{NO}]+[\ce{NO2}]+[\ce{NO2^-}]+[\ce{NO3^-}])$ decreases (Fig.~5b). 
The extrapolated endmember composition with 0\% NO remaining is $\delta ^{15} \ce{N}_0 = -17 \pm 1 \text{\textperthousand}$, the initial $\delta ^{15}$N of the NO reservoir, with an $\epsilon$ factor (i.e. the fractionation between the NO reservoir and the first produced nitrate) of $47 \pm 6 \text{\textperthousand}$. 
We used a linear fit as our experiment can be described by a closed system Rayleigh fractionation where the initial reservoir and the accumulated products remain in contact as supported by further experiments where the flask was let sit idle for several hours after the end of the experiment, leading to even lower values for $\delta ^{15} \ce{N}$ (open squares in Fig.~5a) than in experiments with no additional rest time (circles). 
In natural lightning clouds, we might find some open-system behavior when large rain droplets escape from the system. 
However, the fractionation in an open system is very similar to that of our closed system experiments as we would start with the same $\delta ^{15} \ce{N}_0$ and $\epsilon$ (blue line in Fig.~5b).

Theoretical calculations of equilibrium isotope fractionation have shown that NO gas should be isotopically depleted relative to \ce{N2} and \ce{NO2} \cite{Walters2015}. 
Very light $\delta ^{15}$N values ($-15$ to $-25\text{\textperthousand}$) have also been measured in organic particulates forming in spark discharge experiments \cite{Kuga2014}, consistent with significant isotopic fractionation occurring in lightning chemistry. 
Hence the following sequence of events is the most logical explanation for the trend in our data: 
First, isotopically light NO ($\delta ^{15}$N near $-17\text{\textperthousand}$) is produced in the gas phase.
Our extrapolated endmember for NO ($\delta ^{15} \ce{N}_0 = -17 \pm 1 \text{\textperthousand}$) falls close to the value predicted for the kinetic isotope fractionation in the reaction \ce{N2 + O -> NO + N}, $\epsilon = -17.4 \text{\textperthousand}$\cite{Kuga2014}, resulting from the higher velocity and increased reaction rate of \ce{^{14}N^{14}N} compared to the heavier \ce{^{15}N^{14}N}.
Second, some NO is converted to \ce{NO2} with a positive fractionation, i.e. isotopically heavy NO is preferentially converted to \ce{NO2}. 
Previous studies have reported nitrogen enrichment factors of $29 \text{\textperthousand}$\cite{Li2020} to $36 \text{\textperthousand}$\cite{Walters2016} between these two species, with additional, minor fractionation possible between gaseous and aqueous \ce{NO2}.
Since we find efficient conversion of \ce{NO2} to nitrite and nitrate, the final isotopic composition of these products is very similar to the $\delta ^{15}$N of the \ce{NO2}.
When only a small fraction of NO has been converted (high $f$-value in Fig.~5b), the isotopic composition of nitrate and nitrite falls near $20\text{\textperthousand}$.
Third, progressive conversion of isotopically heavy NO to aqueous species (via \ce{NO2}) renders the residual NO gas phase isotopically lighter. 
Hence, nitrite and nitrate forming at a larger conversion factor (a smaller $f$ in Fig.~5b) become isotopically lighter as well. 

This is supported by our experiments with additional wait time, allowing more NO to convert to aqueous species over time. 
The system trends over time towards an equilibrium with isotopically light aqueous species ($-6$ to $-15\text{\textperthousand}$ and possibly lighter, given more time).
In our experiments, the equilibration time is likely to be affected by diffusion kinetics between the gas and the water reservoir. 
In a natural lightning storm in a water-saturated atmosphere, the equilibration between gaseous NO and aqueous nitrite and nitrate is likely to be reached faster, possibly leading to even lower $\delta ^{15}$N values trending towards the depleted lightning-generated $\delta ^{15}$N(NO). 

\section*{Quantifying lightning as a source of nitrogen for early life}

It has long been postulated that lightning may have driven the origin and early evolution of life on Earth, based on estimates of fluxes and biological demands \cite{Kasting1981,NnaMvondo2001,Navarro-Gonzalez2001,Wong2017}.
In this work we find that nitrogen fixation in an Archean-like (\ce{N2-CO2}) atmosphere is similarly efficient to a modern (\ce{N2-O2}) atmosphere, but further oxidation of NO to nitrite and nitrate is limited by the absence of \ce{O2}. 
The overall production of fixed nitrogen products depends on the Archean lightning flash rate which might have been lower than today's\cite{Wong2017}.
There is still significant uncertainty relating to the \ce{CO2} fraction in the early Archean atmosphere\cite{Catling2020} which will influence the final nitrogen fixation rate\cite{NnaMvondo2001}. 
An additional uncertainty is the total atmospheric pressure in the Archean, where estimates vary between 0.5\cite{Marty2013} and $\SI{3}{\bar}$\cite{Goldblatt2009}.
The velocities of the electrons inside the spark channel, responsible for the dissociation of \ce{N2}, \ce{O2}, and \ce{CO2}, are independent from the gas density, i.e. the pressure \cite{Nijdam2020}.
However, the three-body reactions outside the spark channel depend on the gas number density.
To check whether our results are still valid for different atmospheric pressures, we used simulations with the chemical kinetics network STAND2019 \cite{Rimmer2016,Rimmer2019,Rimmer2019b} to compare the \ce{NO2} production outside the spark channel (at $\SI{300}{\kelvin}$) for pressures of 0.5, 1, and $\SI{3}{\bar}$.
The final \ce{NO2} fraction varied between approximately 80\% (for $\SI{0.5}{\bar}$) and 140\% (for $\SI{3}{\bar}$) compared to $\SI{1}{\bar}$ (our experiments) where we find an \ce{NO2} mixing ratio of $2.8 \times 10^{-5}$.
This uncertainty is much smaller than other factors that influence the \ce{NO2} production rate such as the initial \ce{CO2} fraction in the atmosphere or the lightning flash rate on the early Earth.
A further factor of uncertainty is the availability of \ce{O2} for the production of \ce{NO2} in the Archean atmosphere.
When \ce{O2} is formed by the recombination of atomic oxygen outside of the spark channel, it will be diluted in the background gas.
While our experiments present a limited gas volume in which the chemistry takes place, in real lightning conditions, the \ce{O2} could be transported away from the thunderstorm cloud, reducing the efficiency of \ce{NO2} production. 
We expect the production of \ce{NO2} to happen relatively quickly while the gasses are still in the vicinity of the lightning channel, but kinetic rate modeling is necessary to answer this question definitively.

Our isotopic data allow us to place first empirical constraints on the role of lightning-derived bioavailable nitrogen in the early evolution of life. 
We find the dissolved lightning products significantly lighter ($\delta ^{15} \ce{N} = -6$ to $-15 \text{\textperthousand}$ or less) than the vast majority of sedimentary rocks through geologic time (Fig.~1). 
This conclusion holds when corrected for metamorphic alteration, which is around $1-2 \text{\textperthousand}$ at greenschist facies\cite{Thomazo2013}, applicable to most samples in the compilation.
$\delta ^{15}$N values significantly below $-5 \text{\textperthousand}$ have been documented from one site in the Neoarchean \cite{Yang2019}, interpreted as evidence of high ammonium availability. 
No other settings display such light values, making it unlikely that lightning was a significant source of nitrogen to the biosphere for most of Earth's history. 
It is possible that atmospheric products would have undergone further fractionation after raining out into the ocean, such that the residuum became isotopically heavier and was subsequently trapped in biomass. 
However, such fractionating processes would deplete the nitrite and nitrate reservoir, making it unlikely that sufficient fixed nitrogen remains in solution to support the ecosystem. 
Similarly, UV photolysis can reduce the nitrate and nitrite concentration in the ocean, limiting the contribution to the ecosystem\cite{Ranjan2019}.

Another source of fixed nitrogen on early Earth could have been photochemical production of HCN due to UV irradiation of the atmosphere\cite{Tian2011}.
However, on Titan, where UV radiation and cosmic rays produce a significant amount of HCN, the nitrogen in the HCN is very heavily enriched in \ce{^{15}N} compared to the \ce{N2} ($\delta ^{15}$N values of $\sim 4000\text{\textperthousand}$ (for HCN, ref\cite{vinatier_titan_2007}) and $\sim 650\text{\textperthousand}$ (for \ce{N2}, ref\cite{niemann_composition_2010}), respectively).
If photochemical HCN production had been a significant source of nutrients on the early Earth, we would expect to see a strong enrichment of \ce{^{15}N} in the rock samples.

Our results are thus consistent with the notion that biological \ce{N2} fixation evolved early \cite{Weiss2016}, making the biosphere independent from lightning as a nutrient source. 
There is, however, potentially one exception, that we know of, of rocks from the Paleoarchean, where reconstructed pre-metamorphic $\delta ^{15}$N values may have been as low as $-10\text{\textperthousand}$ (ref\cite{Stueken2021}, pink line in Fig.~1). If correct, those numbers could represent a lightning contribution, meaning that Earth's earliest ecosystems and by extension prebiotic networks may have benefitted from lightning reactions.

To conclude, our results suggest that lightning was not the main source of bioavailable nitrogen for the established biosphere, but it could have been significant for Earth's earliest ecosystems and possibly for life's origins. 
In addition, our results allow the community to investigate the source of fixed nitrogen on other bodies in the Solar System such as Mars where the Mars Science Laboratory and subsequent measurements have detected deposits of nitrate in several locations\cite{Stern2015,Sutter2017}.

\backmatter

\bmhead{Acknowledgments}

We thank Henderson (Jim) Cleaves for technical advice on the experimental setup, Paul Rimmer and the members of the Leverhulme Centre for Life in the Universe (Cambridge) for helpful discussions of our results, and Ben K. D. Pearce for comments on our manuscript.
P.B. acknowledges a St Leonard’s Interdisciplinary Doctoral Scholarship from the University of St Andrews. 
E.E.S. acknowledges funding from a Royal Society research grant (RGS\textbackslash R1\textbackslash 211184) and from a NERC Frontiers grant (NE/V010824/1). 
Ch.H. is part of the CHAMELEON MC ITN EJD which received funding from the European Union’s Horizon 2020 research and innovation programme under the Marie Sklodowska-Curie grant agreement number 860470. 
In order to meet institutional and research funder open access requirements, any accepted manuscript arising shall be open access under a Creative Commons Attribution (CC BY) reuse licence with zero embargo.
This version of the article has been accepted for publication after peer review but is not the Version of Record and does not reflect post-acceptance improvements or any corrections. 
The Version of Record is available online at \url{https://doi.org/10.1038/s41561-023-01187-2}.

\bmhead{Authors' contributions}

E.E.S. and Ch.H. conceived the project; E.E.S. and P.B. built the experimental setup; P.B., L.R. and Y.P. carried out the experiments; P.B., L.R., Y.P. and W.W. performed the analyses; M.C. provided analytical support; P.B., E.E.S., Ch.H. and W.W. analysed the data; P.B. wrote the manuscript with contributions from all authors. 

\bmhead{Competing interests}

The authors declare no competing interests.

\section*{Methods}\label{sec11}

\subsection*{Experimental setup}

All experiments were carried out at the University of St Andrews in the St Andrews Isotope Geochemistry Lab (StAIG). 
We used an experimental setup similar to the one described by Parker et al.\cite{Parker2014} (Fig.~2). 
The spark discharge was contained in the 1-litre reaction flask (Pyrex glass), which contained both the water and the spark electrodes (tungsten metal). 
This flask was connected to a vacuum line (stainless steel) with an ultra-torr fitting and thus could be disconnected at the end of the experiment. 
The sparks were generated by a \textit{BD-50E Heavy Duty Spark Generator} with a maximum voltage of $\SI{49}{\kilo\volt}$. 
Before starting an experiment, the system was evacuated by opening V1 and V3 to the vacuum pump. 
The valve to the reaction flask (V8) was only opened once the rest of the line was evacuated to minimize evaporation of water from the experimental flask. 
Once the pressure within the flask had reached a few $\si{\milli\bar}$, V3 and V1 were closed and the system was filled with \ce{N2} gas via V2 up to $\SI{1}{\bar}$ pressure. 
The \ce{N2} was re-evacuated, and this process was repeated 2 times to completely purge the reaction flask. 
After that, the desired gas mixture was introduced. 
Originally, we aimed for \ce{O2} concentrations of 21\% and 1\% for the modern and low-\ce{O2} experiments, respectively, and a \ce{CO2} concentration of 20\% for the Archean-like experiments. 
However, slow mixing of the gasses in our the set-up led to slightly lower final concentrations of \ce{CO2} and \ce{O2}:
Archean-like: $0.16 \pm 0.03$ \ce{CO2} and $<5\times 10^{-4}$ \ce{O2};
Low-\ce{O2}: $<2\times 10^{-4}$ \ce{CO2} and $(6\pm1)\times10^{-4}$ \ce{O2};
Modern: $<6\times 10^{-4}$ \ce{CO2} and $0.14 \pm 0.03$ \ce{O2}.
The small amounts of \ce{CO2} present in the modern and low-\ce{O2} experiments and of \ce{O2} in the Archean-like experiments is likely due to these gasses being dissolved in the water and not completely removed during evacuation, traces of air remaining in the setup, and recombination of atomic oxygen produced by dissociation of \ce{H2O} and \ce{CO2} in the experiment and the ion source of the mass spectrometer.
For the \ce{O2} in the low-\ce{O2} experiments, we used a flow controller (Bronkhorst EL-FLOW Prestige). 
Before starting the spark discharge, valve V8 was closed to disconnect the reaction flask from the rest of the setup. 
An oxygen sensor was used to monitor the trace amount of oxygen present in anoxic experiments. 
Before and after each experiment, the gas from the flask was allowed to flow towards a quadrupole mass spectrometer gas analyser via V23 (see below for analytical method). 
We did not perform gas analyses during the experiment, because we found that a larger reservoir of NO and \ce{NO2} needed to be generated before reliable measurements could be made. 
When not connected to the experiment, argon gas was fed to the mass spectrometer via V24 and V25 to keep the system free from contaminants. 
After the experiment, the fluid phase was transferred into a 50ml Falcon centrifuge tube for subsequent analyses (see below).

\subsection*{\ce{NO_$x$} gas abundance measurements}
\label{Sec_methods_gas}

Before and after the spark experiment, the gas composition in the flask was analysed with a quadrupole mass spectrometer (Hiden Analytical ExQ Quantitative Gas Analyser) to determine the abundance of all gases, including NO and \ce{NO2}.
The instrument was run in \textit{Multiple Ions Detection} mode to monitor the abundance of several mass/charge ratios (m/z) (12, 14, 15, 16, 17, 18, 20, 28, 30, 32, 40, 46, 48) and thus detect \ce{N2}, \ce{O2}, \ce{CO2}, \ce{NO}, \ce{N2O}, \ce{NO2}, and \ce{O3}.
When analysing the pure \ce{N2}-\ce{O2} gas before running the spark experiment, a peak at m/z 30 was detected. 
This is likely due to the recombination of \ce{^{14}N} and \ce{^{16}O} fragments from the \ce{N2} and \ce{O2}, respectively, to \ce{NO} inside the mass spectrometer \cite{Cardinal2003}. 
By comparing measurements from before and after the spark experiment, this interference could be subtracted.
We found these background measurements to be very stable with variations in the background between 3\% and 17\% for the Archean-like and modern experiments, respectively.
We also measured and subtracted the background for the other masses and used the standard deviation of these background measurements for the error of our gas measurements. 
However, after combining multiple measurements for Fig.~4, the standard deviation of this average is substantially larger than the combined individual errors which scale with $1/\sqrt{N}$ for $N$ data points.

To determine the molecular abundances from the measured intensities, we used the mass spectra provided in the NIST database \cite{NIST}, which accounts for fragmentation of molecules in the ion source of the instrument.
Unfortunately, \ce{N2O} produced signals at m/z 30 and 44, overlapping with peaks for \ce{NO} and \ce{NO2} at m/z 30, and \ce{CO2} at m/z 44. 
However, calculations and experiments suggest that \ce{N2O} production in lightning and discharge experiments is approximately 4 magnitudes lower than \ce{NO} production \cite{Levine1979,Hill1984}, suggesting that the contribution of \ce{N2O} to the peak at m/z 30 can be neglected. 
The abundance of \ce{NO2} could be determined by the signal intensity of the m/z 46 peak; 
the abundance of NO from m/z 30 after subtracting the contribution by the NO-fragment from \ce{NO2}.
For experiments with a high \ce{CO2} abundance, the \ce{CO2} isotopologue \ce{^{16}O^{12}C^{18}O} also contributes to the m/z 46 peak.
We tried to subtract this interference, but the uncertainty is relatively large. 
The data from the \ce{CO2}-rich experiments therefore only gave us an upper limit of the \ce{NO2} concentration (see Fig.~4).
We also found trace amounts of ozone (m/z 48) of $\sim 8 \times 10^{-7} \;\si{bar}$ in the \ce{N2-CO2} experiments and $\sim 4 \times 10^{-5} \;\si{bar}$ in the modern atmosphere experiments, with the detection limit being $\sim 10^{-7} \;\si{bar}$ $(\SI{100}{ppb})$.

\subsection*{Aqueous nitrate and nitrite analyses}

To determine the concentrations of nitrate (\ce{NO3^-}) and nitrite (\ce{NO2^-}) in our solutions, we used an Metrohm 930 \textit{Ion Chromatograph} (IC) with a Metrohm 919 autosampler, a $\SI{150}{\milli\metre}$ Metrosep A Supp 5 separation column ($\SI{4}{\milli\metre}$ bore) with $\SI{3.2}{\milli\molar}$ \ce{Na2CO3} / $\SI{1}{\milli\molar}$ \ce{NaHCO3} anion eluent at a flow rate of $\SI{0.7}{\milli\litre\per\minute}$. 
We ran it using a Metrohm Suppressor Module with a phosphoric and oxalic acid eluent suppressor solution.

The subset of samples shipped to Brown University was analyzed for \ce{NO2^-} and \ce{NO3^-} using colorimetric and IC analytical techniques.
The concentrations of \ce{NO2^-} were determined using a standardized colorimetric technique (e.g., US EPA Method 353.2) involving the diazotization with sulfanilamide dihydrochloride followed by detection of absorbance at $\SI{520}{\nano\metre}$ that was automated using a discrete UV-Vis spectrophotometer (Westco SmartChem).
The analysis of \ce{NO3^-} concentrations was conducted using a reagent-free Dionex Integrion HPIC System with a Dionex AS-HV Autosampler, Dionex AG19 guard and analytical columns, with $\SI{20}{\milli\molar}$ KOH anion eluent at a flow rate of $\SI{1}{\milli\litre\per\minute}$. 

The nitrite and nitrate concentrations in the experiments with different water compositions (see supplementary material) were measured with a colorimetric method using vanadium(III) chloride as a reduction agent \cite{Schnetger2014}.
Analyses were performed with a Thermo Fisher Evolution 220 Computer Control UV-Vis Spectrophotometer with wavelengths set to 530nm for nitrite and 560nm for nitrate. 
This method was used instead of the IC to avoid overloading the IC column with chloride. 

Ammonium concentrations were measured by colorimetry, following the method by Solorzano et al.\cite{Solorzano1969} as described by Cleaves et al.\cite{Cleaves2008}. After mixing with the appropriate amount of reagents, the solutions were analysed at \SI{640}{\nano\metre} with a Thermo Fisher Evolution 220 Computer Control UV-Vis Spectrophotometer.
We found the detection limit of this method to be at $\SI{3.4}{\micro\molar}$.

\subsection*{Nitrogen isotope measurements}
\label{Sec_Methods_NIso}

For the `evaporated residue' method, carried out at the University of St Andrews, $\SI{40}{\milli\litre}$ of each water sample were mixed with $\SI{250}{\milli\gram}$ of \ce{KOH} to form \ce{KNO3} and \ce{KNO2}. 
This solution was then evaporated with a freeze drier for about one week until the residue was completely dry.
The dried residue was weighed into tin capsules and analysed by flash-combustion, using an \textit{elemental analyser (EA IsoLink)} coupled to an \textit{isotope ratio mass spectrometer} (MAT253 Thermo Finnigan) via a ConFlo IV. 
The analyses were calibrated with the international reference materials RM8568/USGS34 ($\delta ^{15} \ce{N} = -1.8 \text{\textperthousand}$; $\delta ^{18} \ce{O} = -28 \text{\textperthousand}$), RM8569/USGS35 ($\delta ^{15} \ce{N} = 2.7 \text{\textperthousand}$; $\delta ^{18} \ce{O} = 57 \text{\textperthousand}$), RM8549/IAEA-NO-3 ($\delta ^{15} \ce{N} = 4.7 \text{\textperthousand}$; $\delta ^{18} \ce{O} = 25.6 \text{\textperthousand}$) \cite{Bohlke1993,Bohlke1995,Bohlke2003}.
We note that adding \ce{KOH} to the sample increases the pH of the solution, which includes small amounts of \ce{NH4^+}. 
Once the pH is increased above 9.25, most \ce{NH4^+} will dissociate to \ce{NH3} which can be lost by volatilisation with a fractionation of $-45\text{\textperthousand}$ \cite{Li2012,Stueken2016}.
Tests to quantify this effect on the bulk measurement of the nitrogen isotope fractionation showed that the ammonium concentration was too small for any volatilisation to affect the measured bulk isotopic value.
A subset of samples was also analyzed for nitrogen and oxygen isotope composition at Brown University using the denitrifer method \cite{Sigman2001,Casciotti2002}.
Briefly, samples were injected into vials containing P. aureofaciens that quantitatively convert nitrate and nitrite to nitrous oxide (\ce{N2O}).
The generated \ce{N2O} was concentrated and purified using an automatic purge and trap system and introduced to a continuous flow isotope ratio mass spectrometer with a modified gas bench interface.
Measurement of \ce{N2O} was conducted at an m/z of 44, 45, and 46 for $\delta ^{15}$N and $\delta ^{18}$O determination, and unknowns were corrected relative to internationally recognized nitrate salt reference materials that included: USGS34, USGS35, and IAEA-NO-3 (see above).
Both methods return the combined isotopic compositions of nitrate and nitrite.

\subsection*{Kinetic chemistry calculations}

To test how the production of \ce{NO2} depends on atmospheric pressure, we performed calculations with the photochemistry and diffusion code ARGO and the chemical kinetics network STAND2019 \cite{Rimmer2016,Rimmer2019,Rimmer2019b}.
STAND2019 contains 3085 forward reactions of 224 charged and 197 neutral species.
For the starting conditions, we assumed that 1\% of the gas mixture was cycled through the spark in our experiments. 
We further assumed that the composition of this portion of the gas corresponds to the chemical equilibrium composition of the initial gas mixture (84\% \ce{N2}, 15\% \ce{CO2}, 1\% \ce{H2O}) at a freeze-out temperature of $\SI{3000}{\kelvin}$.
We calculated the chemical equilibrium composition at $\SI{3000}{\kelvin}$ with GGChem\cite{woitke_equilibrium_2018}, including 87 neutral and charged species containing the elements H, C, N, O, as well as electrons.
We then added the calculated concentrations of H, N, O, \ce{H2}, \ce{N2}, \ce{O2}, OH, NO, CO, \ce{CO2}, \ce{H2O} (in total 1\%) to the background gas (99\%) and used this total gas mixture as input for ARGO/STAND2021.
We performed the kinetic chemistry simulations at a temperature of $\SI{300}{\kelvin}$ and pressures of 0.5, 1, and $\SI{3}{\bar}$ and compared the resulting \ce{NO2} concentrations when steady state was reached after 69, 23, and $\SI{3}{\minute}$ at 0.5, 1, and $\SI{3}{\bar}$, respectively.
Increasing the fraction of gas cycled through the spark channel does not affect the ratio between the \ce{NO2} concentrations for the different gas pressures, but decreases the time scale to reach steady state.

\bmhead{Data availability}

A full methods section, a detailed description of the chemical processes, the description of additional experiments, and a machine-readable table of the data presented in this work is available online.
To access the data please follow \url{https://doi.org/10.5285/81dfa4de-5a47-479f-8de8-15e5ef398072}.

\renewcommand{\thefigure}{S\arabic{figure}}
\renewcommand{\thetable}{S\arabic{table}}

\section*{Supplementary Material}

\subsection*{Chemical pathways}
\label{Sec_Methods_Chemistry}

The production of NO in the spark channel follows the Zel'dovich mechanism \cite{Zeldovich1966} with the following reactions in a \ce{N2-O2} atmosphere\cite{Yung1979,Chameides1986}:
\begin{eqnarray}
    \ce{O2 &<=>& O + O} \\ 
    \ce{O + N2 &<=>& NO + N} \\ 
    \ce{N + O2 &<=>& NO + O}.
\end{eqnarray}
Similar reactions lead to the production of NO in a \ce{N2-CO2} atmosphere\cite{NnaMvondo2001}:
\begin{eqnarray}
    \ce{CO2 &<=>& CO + O} \\ 
    \ce{O + N2 &<=>& NO + N} \\ 
    \ce{N + CO2 &<=>& NO + CO}. 
\end{eqnarray}
In addition, atomic oxygen will recombine to form \ce{O2}.
A schematic showing the chemical reactions in our experiment is shown in Fig.~3.
In both gas compositions, the NO will then oxidise further to \ce{NO2}, which will be in equilibrium with \ce{N2O3} and \ce{N2O4} \cite{Joshi1985,Miller1987}:
\begin{eqnarray}
    \ce{2NO + O2 &->& 2NO2} \\ 
    \ce{NO + NO2 &<=>& N2O3} \\ 
    \ce{2NO2 &<=>& N2O4}.
\end{eqnarray}

Another potential product of further reactions involving NO is \ce{N2O}.
However, discharge experiments \cite{Levine1979} found that the yield of \ce{N2O} is about 4 magnitudes lower than that of \ce{NO} \cite{Levine1981}, $4\times10^{12}\si{molecules\per\joule}$ and $5\times10^{16}\si{molecules\per\joule}$, respectively.
Theoretical calculations \cite{Hill1984} return an even slightly lower \ce{N2O} yield of $8\times10^{11}\si{molecules\per\joule}$.
We therefore neglected any \ce{N2O} production in our analysis.
\ce{NO} can also react to \ce{NO2} in the presence of ozone.
We did measure low ozone concentrations in some of our experiments but it is uncertain how important this pathway is here.
The gas mixture in our experiments contains approximately 1\% of water vapour due to evaporation of the liquid water in the flask at room temperature, which limits the abundance of ozone in the gas mixture\cite{Stark1996}.
In anoxic conditions, Summers et al.\cite{Summers2012} have shown that the disproportionation reaction
\begin{equation}
    \ce{3NO -> N2O + NO2}
\end{equation}
is an important source of \ce{NO2}.
However, in their experiments, \ce{NO} was introduced directly into a gas mixture of \ce{N2} and \ce{CO2}, meaning no spark is present that could provide \ce{O2} from \ce{CO2} and \ce{H2O} dissociation for \ce{NO2} production via Reaction~(7).

\begin{figure}[ht]
    \centering
    \includegraphics[width=0.6\columnwidth]{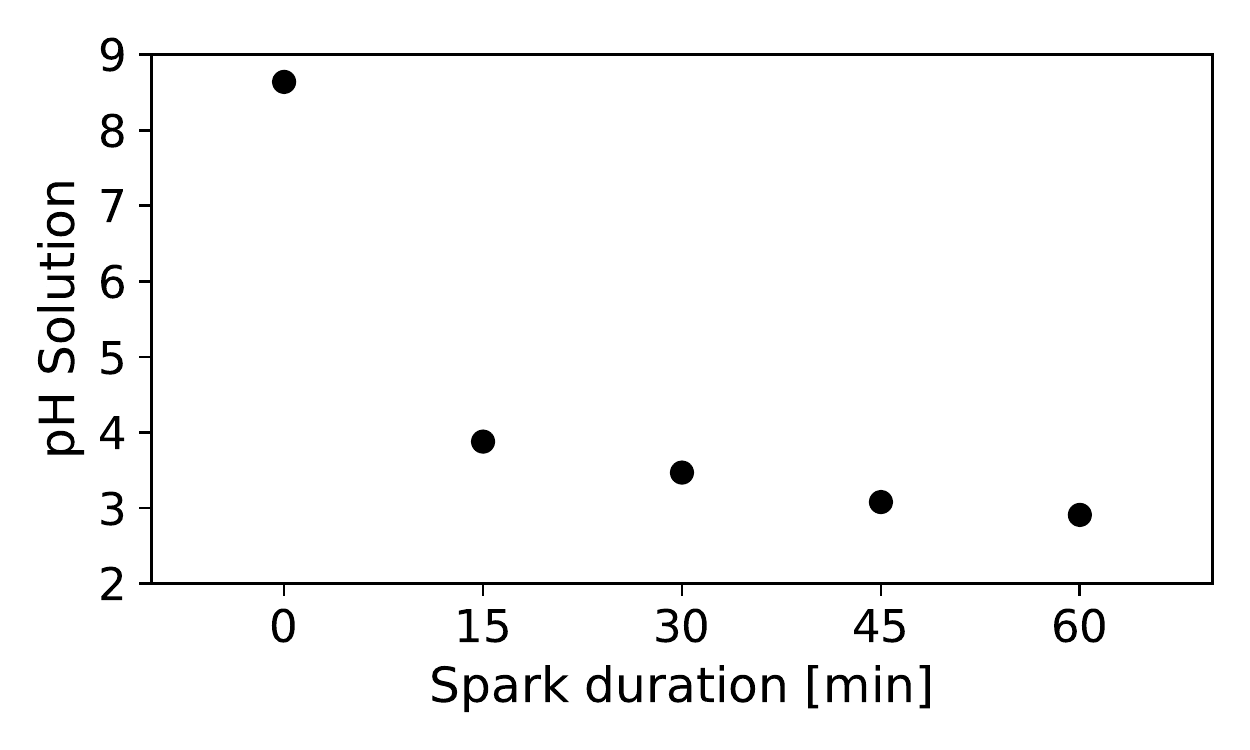}
    \caption{pH of solution at different times during spark experiments. Starting with $\SI{50}{\milli\litre}$ DI water, gas: $\SI{0.9}{\bar}$ \ce{N2} and $\SI{0.1}{\bar}$ \ce{CO2}.}
    \label{Fig_pH}
\end{figure}

\ce{NO2}, \ce{N2O3}, and \ce{N2O4}, will be absorbed into the water (in equilibrium with the partial pressure in the gas) where they will further react with the water to \ce{HNO2} and \ce{HNO3} \cite{Joshi1985, Miller1987}:
\begin{eqnarray}
    \ce{2NO2 + H2O &<=>& HNO2 + HNO3} \label{Reac_NO2_HNO3} \\
    \ce{N2O4 + H2O &<=>& HNO2 + HNO3} \\
    \ce{N2O3 + H2O &<=>& 2HNO2},
\end{eqnarray}
\ce{HNO2} is not stable and will further react to \ce{HNO3},
\begin{equation}
    \ce{3HNO2 <=> HNO3 + H2O + 2NO}.
\end{equation}
\ce{HNO2} and \ce{HNO3} will further react with water to nitrite (\ce{NO2^-}) and nitrate (\ce{NO3^-}), respectively:
\begin{eqnarray}
    \ce{HNO2 + H2O &<=>& NO2^- + H3O^+} \\
    \ce{HNO3 + H2O &<=>& NO3^- + H3O^+}.
\end{eqnarray}
In a low pH environment (Fig.~\ref{Fig_pH}), \ce{HNO2} is more abundant than \ce{NO2^-}, allowing further oxidation to \ce{HNO3} and nitrate.
We found that the nitrite concentrations in our samples were systematically lower if the analyses were performed several days after the experiment rather than on the same day, suggesting that nitrite was continuously converted to nitrate in the aqueous phase.
We speculate that this is due to the presence of another strong oxidizer produced in the spark experiment, such as \ce{H2O2}, which is known to be produced by lightning \cite{Zuo1999}.
Our oxygen isotope data ($\delta ^{18}$O has a consistent value of $20 \; \text{\textperthousand}$ throughout all experiments) further supports this pathway of nitrate formation via nitrite:
In acidic conditions (our solutions have a pH $<4$, see Fig.~\ref{Fig_pH}), nitrite exchanges oxygen with water, completely eradicating the isotopic signature of the initial nitrite \cite{Bunton1959,McIlvin2006}.

The dominant pathway for nitrate formation is indicated in Fig.~3 with bold arrows and goes via Reaction~(10).
Even for our experiments where the final \ce{NO2} pressure is largest, both \ce{N2O3} and \ce{N2O4} pressures are approximately 4 orders of magnitude smaller than the overall \ce{NO_$x$} pressure when using the equilibrium constants \cite{Joshi1985}.
Since the final partial pressure of \ce{NO2} is below $\SI{1}{\milli\bar}$ in all our experiments, \ce{NO2} absorption is the most important route of nitrate formation, as opposed to \ce{N2O4} absorption \cite{Joshi1985,Newman1988}.

\subsection*{Additional experiments}
\label{Sec_Exp_Add}

We tested the influence of the electric field strength on the production of nitrite and nitrate.
The field strength depends on the applied voltage and the size of the gap between the electrodes.
We used an approximation of the electrodes by long and thin ellipsoids \cite{Kohn2015} but could not find a clear correlation to the nitrate production.

\subsubsection*{Spark gap effects}
\label{Sec_Exp_Gap}

\begin{figure}[ht]
    \centering
    \includegraphics[width=0.6\columnwidth]{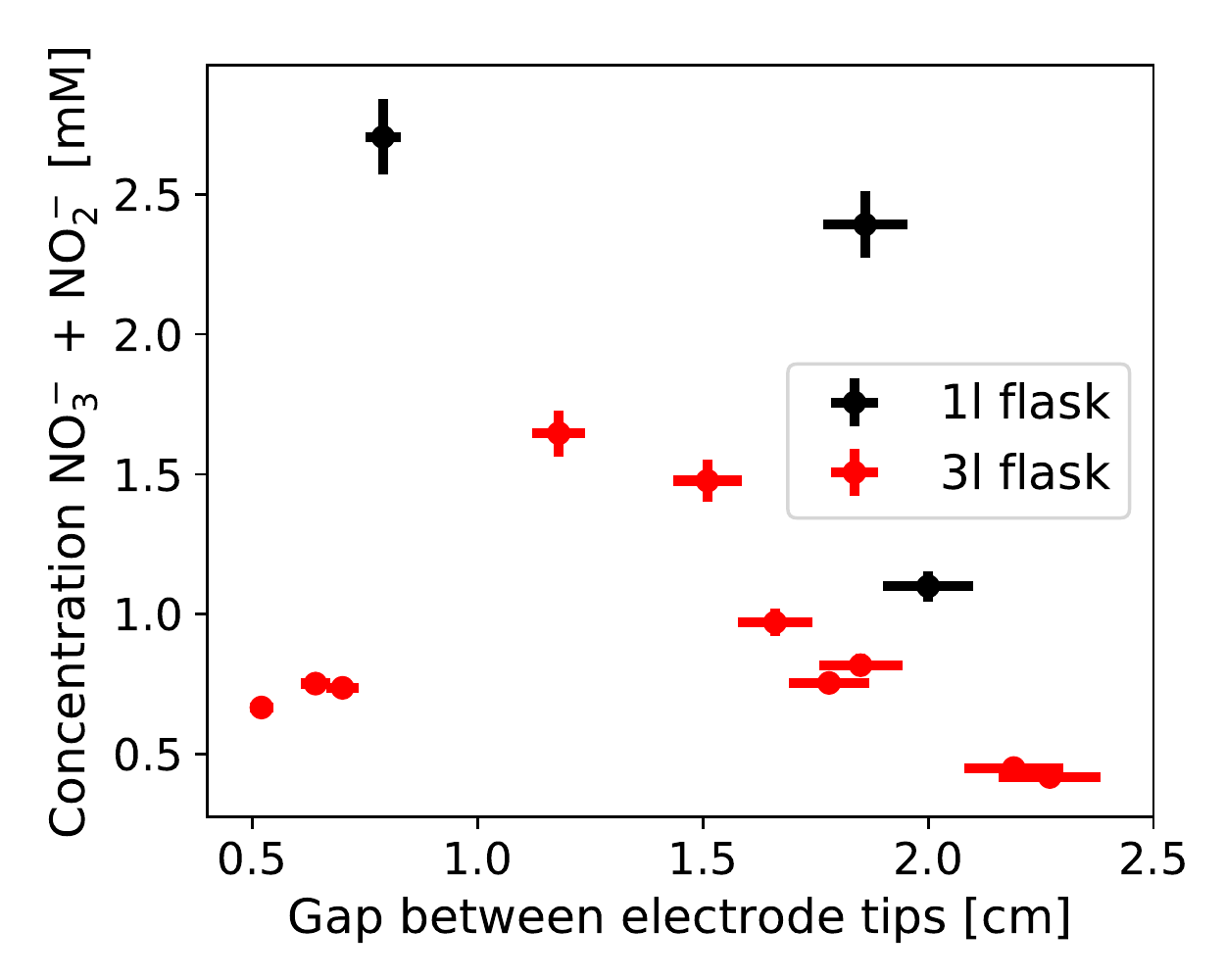}
    \caption{Final Concentration of Nitrate + Nitrite for experiments with varying distance between the electrode tips. Only experiments with modern atmospheric composition. Experiments performed with a $\SI{1}{\litre}$ and $\SI{3}{\litre}$ flask are shown in black and red, respectively. Data points are individual measurements with uncertainties in concentration and distance.}
    \label{Fig_Distance}
\end{figure}

However, experiments with differently sized spark gaps showed an interesting behaviour (Fig.~\ref{Fig_Distance}).
For large spark gaps, the production of nitrate and nitrite dropped as the electric field strength approached the breakdown field in dry air \cite{Kohn2019} of $\SI{35}{\kilo\volt\per\centi\metre}$ and the continuity of the spark was disturbed.
On the other hand, with a decreasing spark gap, the total volume of the spark channel decreased which in return limited the degree of nitrogen fixation, possibly because a larger portion of the energy was lost as heat to the electrodes.
We identified a setting where the nitrate production peaked for spark gaps between $\SI{1}{\centi\metre}$ and $\SI{1.5}{\centi\metre}$ which we used for all further experiments.
Note that some of the initial experiments were carried out with a $\SI{3}{\litre}$ flask, but we transitioned to a $\SI{1}{\litre}$ flask for the majority of the project, because this allowed for a more efficient absorption of the produced nitrogen oxides into the water as the partial pressures of NO and \ce{NO2} were relatively higher in the smaller volume. 
Furthermore, the $\SI{1}{\litre}$ flask was faster to evacuate. 

\subsubsection*{Energy yield}
\label{Sec_Exp_EY}

\begin{figure}[ht]
    \centering
    \includegraphics[width=0.6\columnwidth]{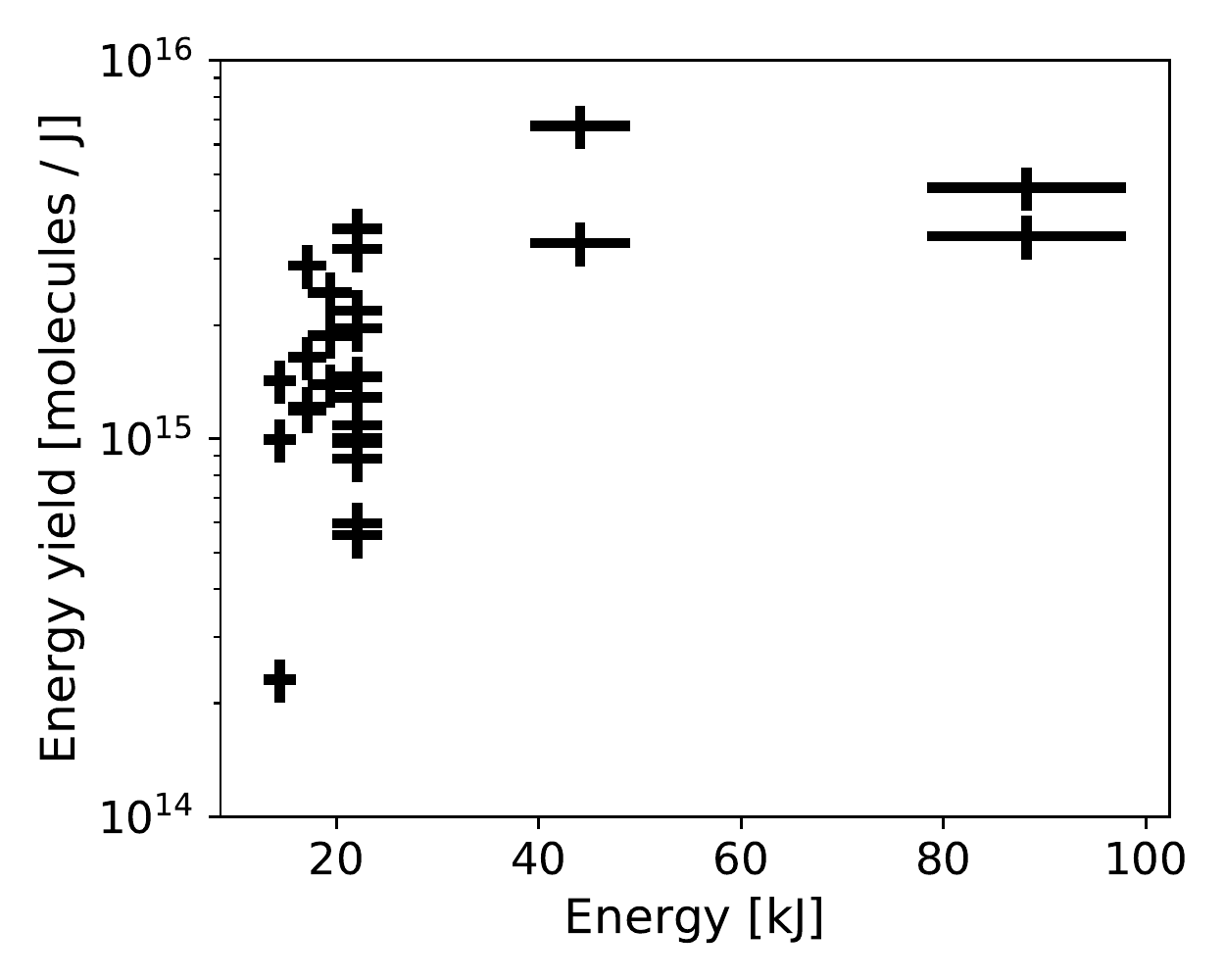}
    \caption{Energy yield (N atoms fixed per Joule) against input energy for production of Nitrate + Nitrite. Only experiments with modern atmospheric composition. Data points are individual measurements with uncertainties in energy and energy yield.}
    \label{Fig_EYield_Energy}
\end{figure}

The rate of nitrogen fixation increases with increasing energy. 
The energy input into the experiment can be calculated with the voltage $U$ and current $I$ of the spark generator and the duration $t$ of the spark: $E = 1/2 \; U I t$. 
The voltage can be changed in increments with $U_{max} = 49 \pm \SI{2}{\kilo\volt}$ and the current is $I = (1 \pm 0.1) \si{\milli\ampere}$.
The duration of the spark ranged between 15 and $\SI{60}{\minute}$.
The energy yield (nitrate and nitrite) increases with increasing energy, levelling off above approximately $\SI{40}{\kilo\joule}$ (Fig.~\ref{Fig_EYield_Energy}).
Reasons for the less efficient production of nitrate and nitrite at lower energies are likely a combination of higher energy losses by heating of the electrodes and less time for further oxidation of \ce{NO_$x$} to nitrate and nitrite as the low-energy experiments typically represent spark durations of $\SI{15}{\minute}$.

\subsubsection*{Effects of duration and water composition}
\label{Sec_Exp_WaterComp}

\begin{figure}[ht]
    \centering
    \includegraphics[width=0.6\columnwidth]{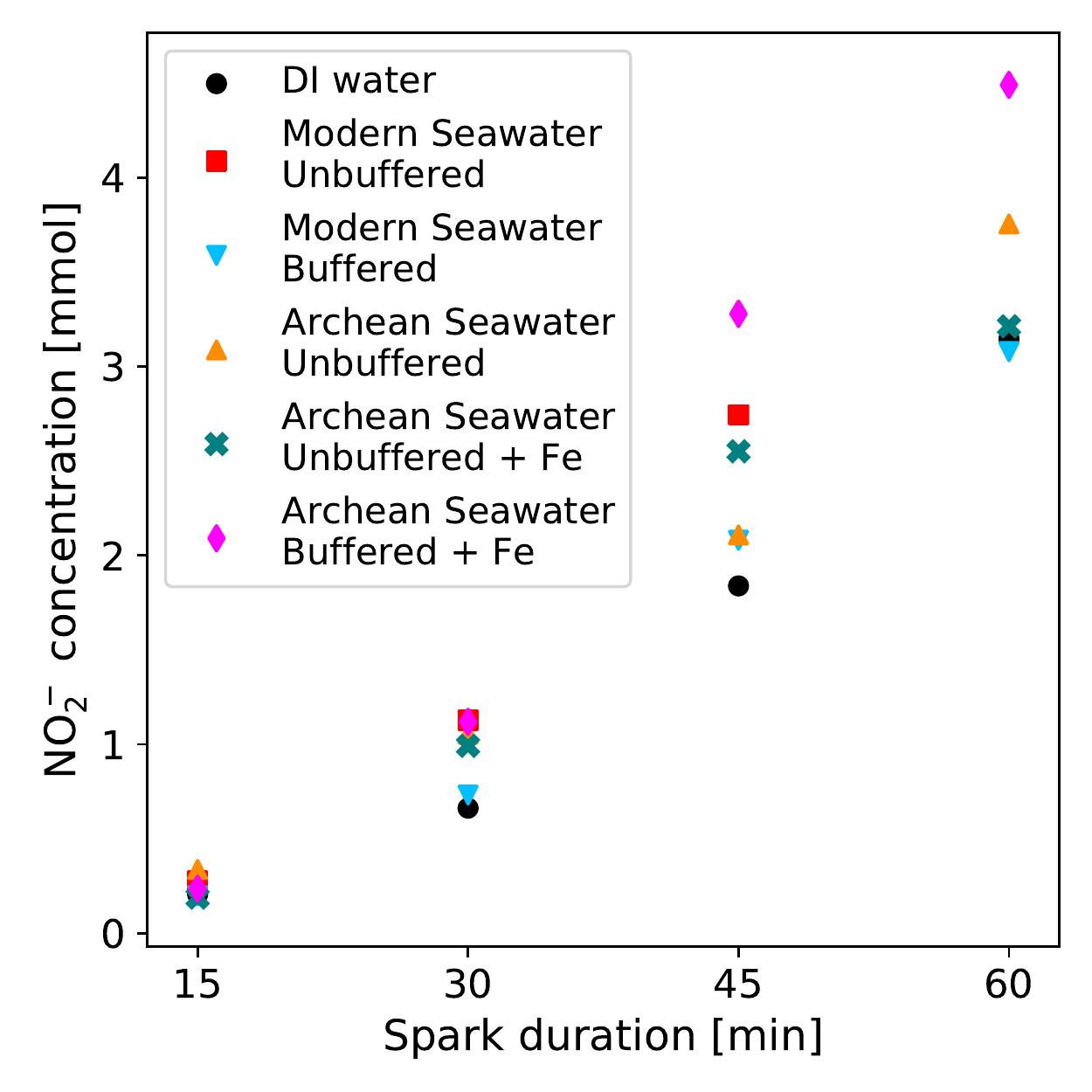}
    \caption{Nitrite (\ce{NO2^-}) concentrations for different durations of spark and different water compositions (Table~\ref{Tab_WaterComp}). Starting with $\SI{50}{\milli\litre}$ of water, gas: $\SI{0.9}{\bar}$ \ce{N2} and $\SI{0.1}{\bar}$ \ce{CO2}.}
    \label{Fig_watercomp}
\end{figure}

\begin{table}[ht]
    \begin{center}
    \caption{Composition of modern and Archean seawater used in experiments \cite{Henson2019} (solvent: DI water). Results presented in Fig.~\ref{Fig_watercomp}.}
    \begin{tabular}{ccc}
    	\toprule
    	Compound & Modern seawater & Archean seawater${}^{a}$ \\ 
    	\midrule
    	NaCl & $\SI{423}{\milli\molar}$ & $\SI{423}{\milli\molar}$ \\
    	\ce{CaCl2} & $\SI{9.27}{\milli\molar}$ & $\SI{9.27}{\milli\molar}$ \\
    	\ce{MgSO4} & $\SI{25.5}{\milli\molar}$ & - \\
    	\ce{NaHCO3} & $\SI{2.14}{\milli\molar}$ & $\SI{2.14}{\milli\molar}$ \\
    	KCl & $\SI{9}{\milli\molar}$ & $\SI{9}{\milli\molar}$ \\
    	\ce{CaCO3}${}^{b}$ & $\SI{1}{\gram}$ & $\SI{1}{\gram}$ \\
    	\ce{FeCl2}${}^{c}$ & $\SI{100}{ppm}$ & $\SI{100}{ppm}$ \\
    	\botrule
    \end{tabular}
    \\
    ${}^{a}$ de-oxygenated by pumping N2 gas through the solution\\
    ${}^{b}$ only for buffered seawater\\
    ${}^{c}$ only for iron containing seawater\\
    \label{Tab_WaterComp}
    \end{center}
\end{table}

A subset of experiments was designed to compare the effect of simulated modern and early Archean seawater on the yields of nitrite, nitrate, and ammonium in the discharge experiments with a constant gas composition of $\SI{0.9}{\bar}$ \ce{N2} and $\SI{0.1}{\bar}$ \ce{CO2} to simulate an anoxic neutral Earth's early atmosphere. 
Four different liquid compositions \cite{Henson2019} were used:
\textit{(1)} DI water
\textit{(2)} Modern seawater
\textit{(3)} Modern seawater with \ce{CaCO3} buffer
\textit{(4)} Archean seawater.
First, $\SI{500}{\milli\litre}$ of the solution were prepared in a volumetric flask (see Table~\ref{Tab_WaterComp}) and the initial pH was measured. 
Then $\SI{50}{\milli\litre}$ of the solution were transferred to the reaction flask of the spark discharge experiment, using a syringe. 
The flask was evacuated and flooded with \ce{N2} gas three times before $\SI{0.9}{\bar}$ \ce{N2} and $\SI{0.1}{\bar}$ \ce{CO2} were added. 
Finally, the spark generator was turned on. 
For each liquid composition, four tests were done with discharge times of 15, 30, 45, and $\SI{60}{\minute}$. 
After the experiment, the liquid in the flask was collected and the yield of nitrite, nitrate, and ammonium were analysed with the colorimetric methods (see above).

The yield of nitrite showed a linear increase with discharge time, from ca. $\SI{0.2}{\milli\molar}$ after $\SI{15}{\minute}$ to ca. $\SI{3.5}{\milli\molar}$ after $\SI{60}{\minute}$ of sparking (Fig.~\ref{Fig_watercomp}). 
The minimum yield was found in the experiment with pure water while that the maximum yield appeared with buffered Archean seawater and iron (for the $\SI{60}{\minute}$ experiment). 
However, the difference among those results is not significant. 
The nitrate yields, although showing a general increasing trend with reaction time (not shown), are more irregular, likely due to variable degrees of nitrite conversion to nitrate.
Ammonium was again several orders of magnitude below nitrite and nitrate concentrations even in the presence of ferrous iron, which can theoretically reduce nitrite and nitrate to ammonium\cite{Summers1993}.


\end{document}